%% file: main.tex
\documentclass[10pt,twocolumn,letterpaper]{article}
\usepackage[pagenumbers]{iccv} % To force page numbers, 、
\usepackage{graphicx}
\usepackage{caption}
\usepackage{lipsum}
\usepackage[table,dvipsnames]{xcolor}
\usepackage{adjustbox}
\usepackage{pifont}
\usepackage{amssymb}     
\usepackage{colortbl}       
\definecolor{mygray}{RGB}{240,240,240}
\usepackage{array}
\usepackage{multirow}
\usepackage{booktabs} 
\usepackage{multirow}
\newcommand{\cmark}{\ding{51}}   % 定义对号
\usepackage{makecell}
\definecolor{lightblue}{RGB}{217,235,255}  % 更浅的蓝色

\input{preamble}

\definecolor{iccvblue}{rgb}{0.21,0.49,0.74}
\usepackage[pagebackref,breaklinks,colorlinks,allcolors=iccvblue]{hyperref}

\def\paperID{3269} % *** Enter the Paper ID here
\def\confName{ICCV}
\def\confYear{2025}

\title{Edit360: 2D Image Edits to 3D Assets from Any Angle}

\author{Junchao Huang$^{1}$ \quad Xinting Hu$^{2}$ \quad Shaoshuai Shi$^{4}$ \quad Zhuotao Tian$^{3}$ \quad Li Jiang$^{1,\dagger}$\\
\normalsize$^1$The Chinese University of Hong Kong, Shenzhen \qquad 
\normalsize$^2$Nanyang Technological University\\
\normalsize$^3$Harbin Institute of Technology, Shenzhen \qquad 
\normalsize$^4$Voyager Research, Didi Chuxing\\
}

\begin{document}
\twocolumn[{
\renewcommand\twocolumn[1][]{#1}
\maketitle
\begin{center}
    \captionsetup{type=figure}
    \includegraphics[width=\textwidth]{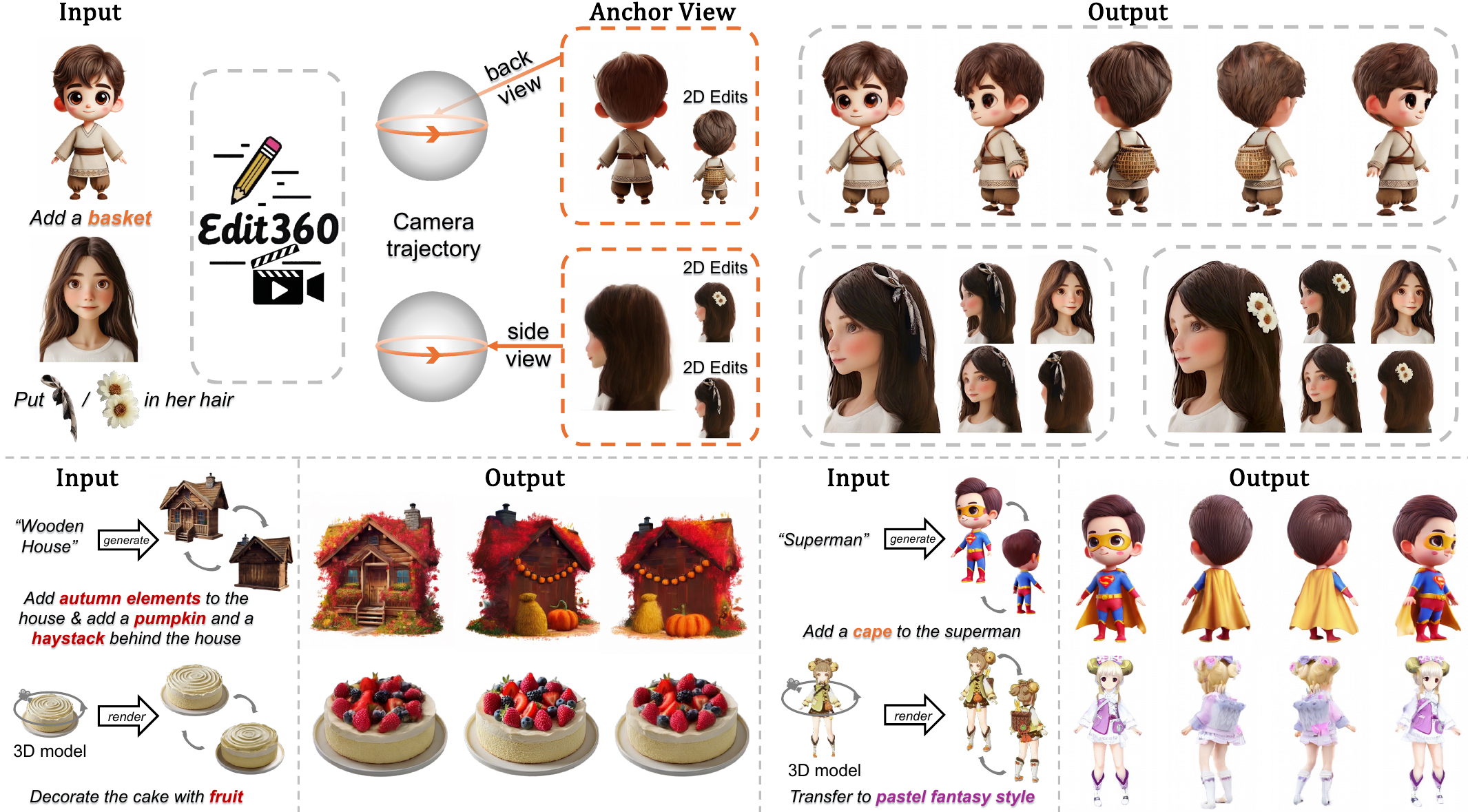}
    \vspace{-5mm}
    \caption{Edit360 enables propagation of 2D edits from any viewpoint to 3D assets, preserving geometric consistency and visual coherence across all views. The framework supports various user inputs, including text descriptions, reference images, and existing 3D assets, allowing for precise and customizable 3D editing. Project page at \href{https://junchao-cs.github.io/Edit360-demo/}{https://junchao-cs.github.io/Edit360-demo/}.}
    \vspace{0mm}
    \label{fig:Teaser}
\end{center}
}]

\renewcommand*{\thefootnote}{$\dagger$}
\footnotetext[1]{Corresponding authors.}

\input{sec/0_abstract}    
\input{sec/1_intro}

\input{sec/2_relatedwork}
\input{sec/3_method}

\input{sec/4_experiments}

\input{sec/5_conclusion}

{
    \small
    \bibliographystyle{ieeenat_fullname}
    \bibliography{main}
}

\end{document}

%% file: preamble.tex
\usepackage{float}

%% file: sec/0_abstract.tex
\begin{abstract}
Recent advances in diffusion models have significantly improved image generation and editing, but extending these capabilities to 3D assets remains challenging, especially for fine-grained edits that require multi-view consistency. Existing methods typically restrict editing to predetermined viewing angles, severely limiting their flexibility and practical applications.
We introduce Edit360, a tuning-free framework that extends 2D modifications to multi-view consistent 3D editing. Built upon video diffusion models, Edit360 enables user-specific editing from arbitrary viewpoints while ensuring structural coherence across all views. The framework selects anchor views for 2D modifications and propagates edits across the entire 360-degree range. To achieve this, Edit360 introduces a novel Anchor-View Editing Propagation mechanism, which effectively aligns and merges multi-view information within the latent and attention spaces of diffusion models. The resulting edited multi-view sequences facilitate the reconstruction of high-quality 3D assets, enabling customizable 3D content creation.
\end{abstract}

%% file: sec/1_intro.tex
\section{Introduction}
\label{sec:intro}
Diffusion models~\cite{Ho_Jain_Abbeel_2020,song2020score} have revolutionized image generation~\cite{Rombach_Blattmann_Lorenz_Esser_Ommer_2022,saharia2022photorealistic} and editing~\cite{10377881}, enabling high-quality creations based on textual descriptions or image inputs, detailed local edits~\cite{chen2024anydoor}, and style transformations~\cite{kotovenko2019content,ruiz2023dreambooth,wang2024instantstyle}. However, extending these capabilities to 3D remains challenging, particularly for fine-grained edits that require consistent propagation of edits across all viewpoints.

Recent advancements in 3D asset generation leverage diffusion models to enhance visual quality, significantly improving texture details and realism~\cite{poole2022dreamfusion,wang2024prolificdreamer,lin2023magic3d,metzer2023latent,watson2022novel,liu2023zero,Liu2023SyncDreamerGM,mercier2024hexagen3d,deitke2024objaverse,tang2024mvdiffusion++}.
Score Distillation Sampling (SDS)-based methods~\cite{poole2022dreamfusion,wang2024prolificdreamer,lin2023magic3d} optimize 3D representations with 2D priors but often suffer from slow convergence, view inconsistency, and the ``Janus problem".
Fine-tuning approaches\cite{liu2023zero,Liu2023SyncDreamerGM,mercier2024hexagen3d} improve cross-view coherence but remain limited by sparse-view training, making consistent edits propagation across multiple views challenging.

In contrast, Video 3D Diffusion Models (V3DMs)~\cite{voleti2024sv3d,chen2024v3d} generate consistent dense view sequences following the pre-defined camera trajectory from a single front-view image. This property makes them well-suited for editing tasks, as modifications made to one view can be naturally propagated across others.  
However, current V3DMs are constrained to single front-view inputs, allowing edits only on the front view before generation. This limitation prevents modifications of occluded or less-visible regions.
Directly editing alternative viewpoints (e.g., the side/back view) and replacing the original input view results in the loss of identity information provided by the front view.
This challenge motivates the development of an approach that enables precise edits tailored to any chosen viewpoint, while maintaining global structural coherence and original identity information.

We introduce Edit360, a tuning-free framework for multi-view consistent 3D editing based on V3DMs. Unlike prior methods constrained to front-view modifications, Edit360 enables user-specified edits from any viewpoint while preserving the asset’s global identity and structure. Given a user-provided editing instruction, Edit360 selects anchor views, which are the most suitable perspectives for a given edit. After performing precise 2D modifications on these anchor views using off-the-shelf image editing models~\cite{10377881,wang2024instantstyle}, Edit360 propagates these edits across all viewpoints through the proposed Anchor-View Editing Propagation mechanism. This approach ensures that local edits remain structurally and semantically coherent across the entire 360-degree range while preserving the identity of the object in the reconstructed 3D asset.

Anchor-View Editing Propagation extends current V3DMs by introducing an additional camera trajectory starting from the anchor view. This propagation mechanism consists of two complementary components: 
Spatial Progressive Fusion (SPF) and Cross-View Alignment (CVA). SPF aligns these two trajectories through circular-shift operations and progressively merges them using proximity-based weighting, while CVA enforces structural coherence by injecting structural priors through cross-view attention mechanisms to prevent artifacts when conflicting details arise. 
This integrated approach enables Edit360 to preserve user-intended edits specified at arbitrary viewing angles, while maintaining global 3D consistency.

We evaluate Edit360 through qualitative and quantitative analyses. To quantitatively assess multi-view consistency, we integrate an unedited anchor view into the novel-view synthesis process, verifying that its inclusion does not degrade the structural integrity of the 3D model. This confirms that Edit360 seamlessly incorporates multiple anchor views without disrupting the underlying 3D representation.
Qualitative results demonstrate that edits applied to the anchor view propagate effectively across all viewpoints, ensuring spatially consistent modifications. These findings validate Edit360’s ability to propagate 2D edits into fully coherent 3D assets while achieving high generation quality.

Overall, our contributions can be summarized as follows: 
\begin{itemize}
\item We introduce Edit360, a tuning-free framework for multi-view consistent 3D editing, enabling fine-grained modifications from any viewing angle of 3D assets.
\item We propose Anchor-View Editing Propagation with Spatial Progressive Fusion (SPF) and Cross-View Alignment (CVA) components, enabling seamless propagation of edits across dense viewpoints with structural coherence.
\item Comprehensive experiments demonstrate that Edit360 achieves state-of-the-art performance across diverse tasks, including precise 3D editing from multiple viewpoints, global style transformation, and multi-view conditional generation.
\end{itemize}

%% file: sec/2_relatedwork.tex
\section{Related Works}
\label{sec:formatting}
\textbf{Image Diffusion Model for 3D Generation.}
The task of 3D object generation has long been limited by the scarcity of large-scale 3D datasets, limiting content quality and diversity. DreamFusion~\cite{poole2022dreamfusion} addressed this by introducing Score Distillation Sampling (SDS), optimizing 3D representations using pre-trained 2D diffusion models. By leveraging rich priors from large-scale 2D image models, DreamFusion achieved high-quality 3D content without extensive 3D data. However, SDS-based approaches~\cite{poole2022dreamfusion,wang2024prolificdreamer,lin2023magic3d,metzer2023latent,watson2022novel} face limitations such as a front-view bias, causing the “multi-face Janus problem”. To overcome this, Zero123~\cite{liu2023zero} fine-tuned image diffusion models with 3D data, enabling novel views from a single image. However, despite the 3D supervision introduced by Zero123~\cite{liu2023zero} and its subsequent methods~\cite{Liu2023SyncDreamerGM,mercier2024hexagen3d,deitke2024objaverse}, image diffusion-based models still struggle with multiview consistency and can only generate sparse views due to capacity and computational constraints,
making it challenging to directly use these views for 3D reconstruction.

\vspace{1mm}
\noindent
\textbf{Video Diffusion Model for 3D Generation.}
To mitigate these challenges, video diffusion models~\cite{blattmann2023stable,voleti2022mcvd,girdhar2023emu} have emerged as a powerful alternative for generating dense, continuous sequences of frames with high inter-frame consistency.
ViVid-1-to-3~\cite{kwak2024vivid} demonstrated that integrating a pre-trained video diffusion model with multiview generation significantly enhances the consistency of the generated views. In a similar vein, Envision3D~\cite{pang2024envision3d} proposed a two-stage pipeline where a multiview generation model produces sparse anchor views, which are then interpolated by a fine-tuned video diffusion model to generate dense, consistent views. Models like SV3D~\cite{voleti2024sv3d}, V3D~\cite{chen2024v3d}, and VFusion3D~\cite{han2025vfusion3d} directly fine-tuned video diffusion models to generate smooth 360-degree orbit videos, offering consistent multi-view generation from a single input. 

\vspace{1mm}
\noindent
\textbf{3D Object Editing.}
Traditional 3D editing methods, including explicit geometric deformations~\cite{yuan2021revisit,sorkine2005laplacian,sorkine2007rigid,jacobson2012fast,magnenat1988joint,sederberg1986free,yifan2020neural,sumner2005mesh,gao2016efficient,gao2019sparse}, implicit representations~\cite{liu2019neuroskinning,tan2018variational,xu2020rignetneuralriggingarticulated} like NeRFs~\cite{yuan2022nerf,yang2021learning,liu2021editing}, and hybrid approaches~\cite{chen2019learningimplicitfieldsgenerative,park2019deepsdflearningcontinuoussigned} with data-driven priors, are typically designed for specific objects and scenes, resulting in limited generalization capabilities~\cite{qi2024tailor3d}.
These challenges have led to a shift toward using advanced 2D image editing techniques for more intuitive and adaptable 3D customization.
Methods like Control3D~\cite{chen2023control3dcontrollabletextto3dgeneration} and Generic 3D~\cite{chen2024generic3ddiffusionadapter} enable global control in 3D generation by injecting text or image conditions into a 2D diffusion model, then transferring these conditions across multiple views.
While effective for maintaining overall consistency, these methods lack precision for localized edits. 
Recent work, such as Tailor3D~\cite{qi2024tailor3d}, enables the fusion of front and back view 2D local edits into 3D representations, but remain constrained by predetermined viewing angles and limited geometric and texture details.

%% file: sec/3_method.tex
\begin{figure}
  \centering
  \includegraphics[width=\linewidth]{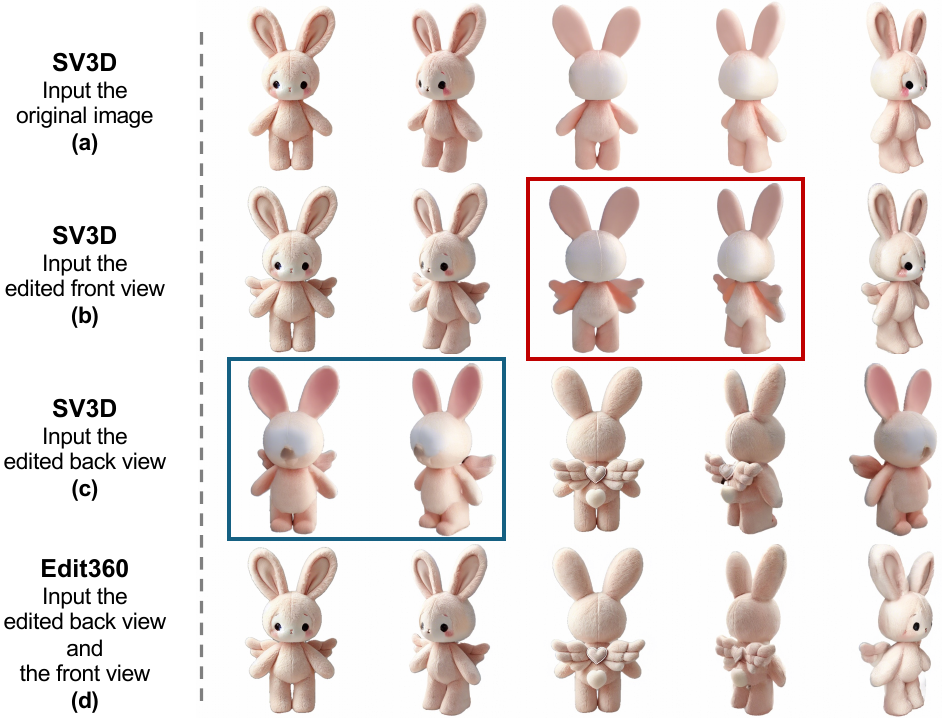}
  \vspace{-6mm}
  \caption{Example of adding wings to a rabbit doll demonstrating the limitations of existing single-view input V3DMs methods (\eg, SV3D~\cite{voleti2024sv3d}) for 3D editing. With only the edited front view as input, SV3D generates incomplete wings that appear fused with the arms (red box). Conversely, using the edited back view as input results in loss of facial identity information (blue box). Our Edit360 approach integrates multi-view information to ensure consistent and complete edits across the entire 3D representation.}
  \vspace{-4mm}
  \label{fig:V3DM}
\end{figure}

\begin{figure*}
\centering
\includegraphics[width=\textwidth]{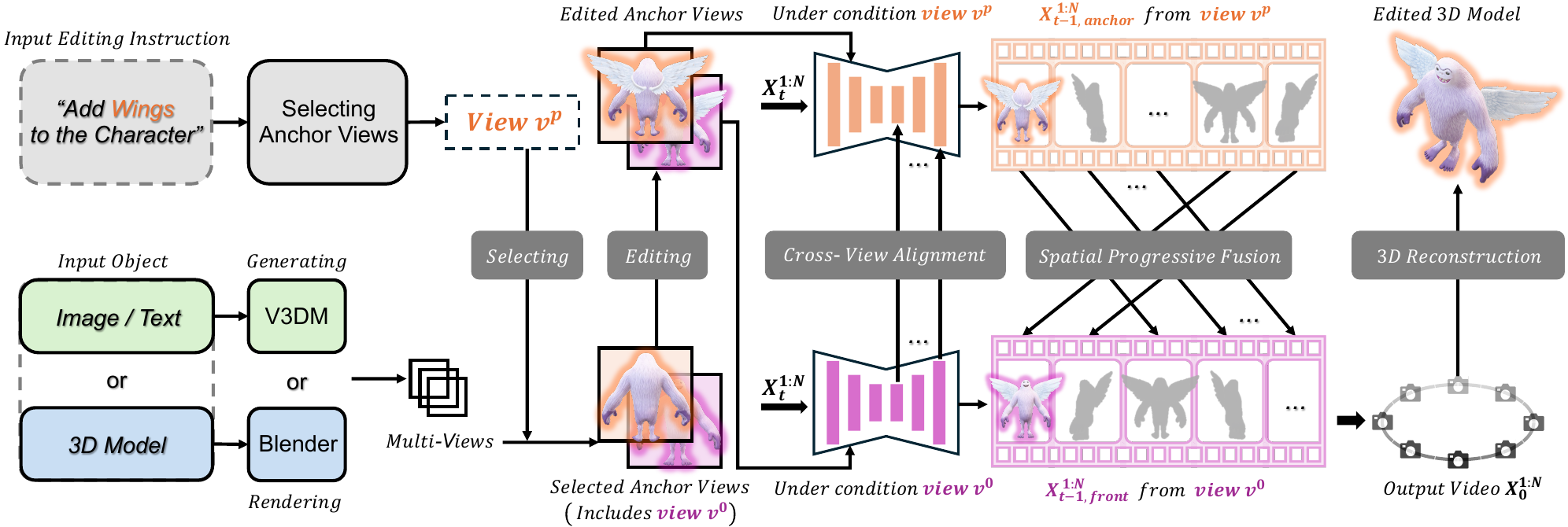}
\vspace{-5mm}
\caption{
{Overview of the Edit360, from input instruction and object (text, image, or 3D model) to the edited 3D asset. With $v^p$ (the back view for this case) selected as the anchor view for editing while $v^0$ (front view) serving to preserving identity information, our Dual-Stream Diffusion Network progressively generates and fuses multi-view sequences through Spatial Progressive Fusion (SPF) and Cross-View Alignment (CVA) at each sampling step, ensuring consistent view generation for reconstructing the edited 3D asset.}
}
\vspace{-4mm}
\label{fig:pipeline}
\end{figure*}

\section{Overall Framework}
\label{sec:Methodology}
This section presents the overall framework of Edit360, a novel approach for user-guided 3D asset editing that requires no training or fine-tuning. Section \ref{subsec:Preliminaries} introduces background on video diffusion models and dense-view synthesis. Building on these models, Section \ref{subsec:Pipeline} details the Edit360 pipeline. Additionally, we discuss how Edit360 enables edits from any viewpoint.

\subsection{Preliminaries}
\label{subsec:Preliminaries}
% During the video generation process, 
\textbf{Video Diffusion Models (VDMs)} generate a video of $N$ frames, $X^{1:N}$, by progressively denoising an initial Gaussian noise sequence over $T$ timesteps. At each timestep $t$, the VDM network $\theta$ predicts noise $\epsilon_{\theta}(X^{1:N}_{t}, t, C)$, where $C$ represents conditioning  text or image inputs. The iterative denoising process follows:
\begin{equation}\label{eq:1} 
X^{1:N}_{t-1} = \text{Denoise}(X^{1:N}_{t}, \epsilon_{\theta}(X^{1:N}_{t}, t, C)),
\end{equation}
where the final output $X^{1:N}_0$ is the generated video $X^{1:N}$. 

\vspace{1mm}
\noindent
\textbf{Video 3D Diffusion Models (V3DMs)}~\cite{voleti2024sv3d, chen2024v3d} extend VDMs to generate 3D-consistent dense-view sequences by fine-tuning them on novel-view synthesis tasks. 
Instead of conditioning on the general text or image, V3DMs take a front-view image $v^0$ as $C$ and predict the sequence of views denoted as: $\{v^0, \dots, v^{N-1}\}$. This sequence represents a full 360-degree orbit around the object along a predefined camera trajectory, where each view $v^p$ corresponds to a view rotated clockwise by $\frac{360}{N} \times p$ degrees relative to $v^0$.

\subsection{The Pipeline of Edit360}
\label{subsec:Pipeline}
While dense-view synthesis based on Video Diffusion Models (VDMs) can generate high-quality multi-view sequences, their dependence on single-view inputs constrains editing flexibility. For instance, when target edits (such as wings) have limited visibility in the front view, texture and shape inconsistencies emerge in subsequently generated views (Figure~\ref{fig:V3DM} (b)). Conversely, performing edits directly in the optimal view for a specific modification (such as the back view where the wing area has maximum visibility) and merely substituting the input view with this edited view results in identity information loss in the reconstructed 3D asset (Figure~\ref{fig:V3DM} (c)).

To address these limitations, Edit360 introduces an additional editable anchor view, the optimal view with maximum visibility for the targeted edit. This anchor view can be either manually specified by users or automatically identified using a Large Language Model (LLM). After editing the anchor view, Edit360 employs Anchor-View Editing Propagation mechanism to synchronize the 2D edits across all views. The mechanism consists of Spatial Progressive Fusion (SPF) and Cross-View Alignment (CVA) to achieve view-consistent editing while maintaining the identity information from the front view. Details are provided in Section~\ref{subsec:Anchor-View Editing Propagation}. This process generates an edited dense-view sequence, which is subsequently used for reconstructing the final edited 3D asset through NeuS~\cite{Wang_Liu_Liu_Theobalt_Komura_Wang_2021} or 3DGS~\cite{kerbl20233dgaussiansplattingrealtime}. As shown in the bottom row of Figure~\ref{fig:V3DM}, Edit360 successfully adds geometrically consistent, richly textured wings to the rabbit doll while preserving its facial characteristics. The overall workflow of Edit360 is illustrated in Figure~\ref{fig:pipeline}.

\vspace{1mm}
\noindent
\textbf{Multi-Format Input Processing.} 
Edit360 supports diverse formats for editable objects, including text, image, and existing 3D assets. All inputs are standardized into multi-view sequences through format-specific methods:
(1) For text inputs, we first synthesize a front view via a text-to-image model~\cite{10140348} then generate multi-views using V3DM~\cite{voleti2024sv3d, chen2024v3d};
(2) For image inputs, we directly apply V3DM for view synthesis;
(3) For existing 3D assets, we use Blender for multi-view rendering.

\section{Anchor-View Editing Propagation}
\label{subsec:Anchor-View Editing Propagation}
To ensure that edits made on the anchor view propagate consistently across the entire dense-view sequence, Edit360 introduces Anchor-View Editing Propagation, a mechanism designed to synchronize modifications across multiple perspectives while preserving global 3D coherence.

In standard V3DMs, the \textit{\textbf{front-view camera trajectory}} is used to generate a dense sequence of views by progressively rotating the camera around the object. The trajectory starts from the front view $v^0$ and follows a clockwise orbit:
\begin{equation}\label{eq:2} 
X^{1:N}_\text{front} = \{v^0, v^1, \dots, v^{N-1}\},
\end{equation}
where each view $v^p$ is rotated by $\frac{360}{N} \times p$ degrees relative to the front view.
To enable edits from arbitrary angles, Edit360 first introduces an \textit{\textbf{anchor-view camera trajectory}}, which follows the same camera rotation principle but originates from the user-selected anchor view $v^p$:
\begin{equation}\label{eq:3} 
X^{1:N}_\text{anchor} = \{v^p, v^{p+1}, \dots, v^{(p+N-1) \mod N}\}.
\end{equation}
Unlike $X^{1:N}_\text{front}$ that remains unedited, $X^{1:N}_\text{anchor}$ is generated after modifying the anchor view, ensuring that edits at $v^p$ influence surrounding views.

Anchor-View Editing Propagation aims to progressively fuse these two trajectories, ensuring seamless integration of anchor-view modifications while preserving 3D priors from the front-view trajectory.  To achieve this, Edit360 integrates Spatial Progressive Fusion (SPF) to blend the two trajectories through spatial alignment and progressive fusion strategies. Additionally, Cross-View Alignment (CVA) resolves structural inconsistencies between them by enforcing feature-level consistency across corresponding views. These two components work together to perfectly merge the otherwise inconsistent view sequences resulting from anchor-view edits, ensuring that modifications are smoothly propagated across the entire 360-degree sequence while maintaining structural integrity and creating a coherent 3D representation.

\subsection{Spatial Progressive Fusion}
\label{subsec:SPF}
Spatial Progressive Fusion (SPF) aligns and blends the edited dense views $X^{1:N}_{\text{anchor}}$ along anchor-view camera trajectory with the original $X^{1:N}_{\text{front}}$ along front-view camera trajectory, ensuring smooth propagation of edits while maintaining spatial consistency.

Since each trajectory originates from a different viewpoint, SPF first aligns the indexed view location of $X^{1:N}_{\text{anchor}}$ to $X^{1:N}_{\text{front}}$. This is achieved using a circular-shift (CS) operation:
\begin{equation}\label{eq:4} 
\text{CS}(X_{\text{anchor}}^{i}) = X_{\text{anchor}}^{(i+N-p)\ \text{mod}\ N}, \quad i=0,1,\dots,N.
\end{equation}
This alignment ensures that each indexed view $i$ in $\text{CS}(X_{\text{anchor}}^{i})$ correspond to the view at the same camera location, where $p$ represents the index of the anchor view.

After alignment, the dense views along two trajectories are merged during each denoising timestep $t\in \{0,1,\dots T\}$:
\begin{equation}\label{eq:5} 
X_{t,\text{SPF}}^i = (1-\alpha^i) \cdot X_{t, \text{front}}^i + \alpha^i \cdot \text{CS}(X_{t, \text{anchor}}^i),
\end{equation}
where the spatial weight $\alpha^i$ dynamically adjusts the contribution of views from each trajectory based on its proximity to the anchor view. $\alpha^i$ decreases as the cyclic distance between view $v_i$ and the anchor view $v_p$ increases, prioritizing edits in spatial proximity to the anchor view. 
% In practice,  we implement  $\alpha^i$ with a Gaussian function, with details provided in the supplementary material.

In later diffusion stages, SPF integrates edge~\cite{996} and texture~\cite{Fogel1989GaborFA} information to refine structural details and mitigate over-smoothing, ensuring fine-grained consistency across views. 
% A detailed breakdown of this multi-scale fusion technique is provided in the supplementary material.

By progressively interpolating between trajectories over multiple timesteps, SPF guarantees that edits introduced in $X^{1:N}_{\text{anchor}}$ are smoothly integrated into $X^{1:N}_{\text{front}}$, maintaining both edit consistency and overall spatial coherence.

\subsection{Cross-View Alignment}
\begin{figure}
\centering
\includegraphics[width=0.99\linewidth]{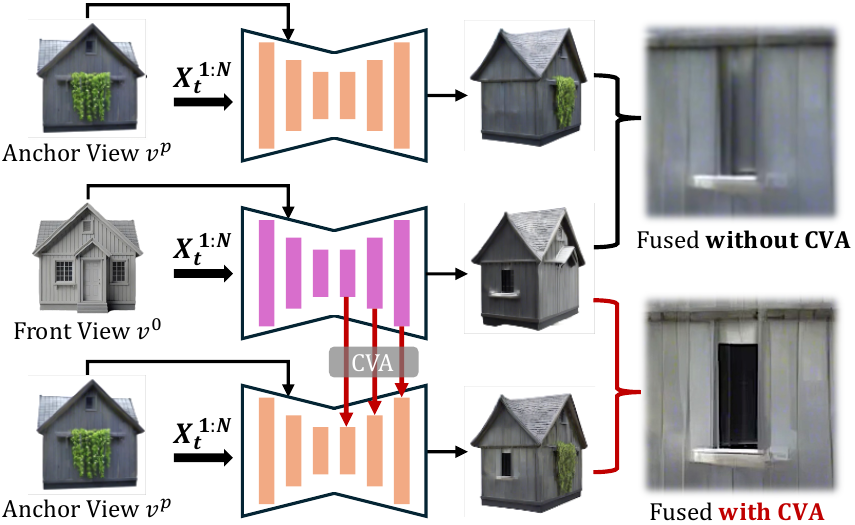}
\vspace{-1.5mm}
\caption{Comparison of fusion results with and without Cross-View Alignment (CVA). The top row shows the result without CVA, where the window is blurred, while the bottom row shows the result with CVA, preserving a clear window in the side view.}
\vspace{-3.5mm}
\label{fig:CVA}
\end{figure}
While Spatial Progressive Fusion (SPF) ensures smooth spatial blending of the front-view and anchor-view trajectories, inconsistencies may still arise due to the inherent randomness of the denoising process. For example, as shown in Figure~\ref{fig:CVA}, a side view of a house generated from the front view includes windows, while the same side view generated from the back view lack them. Directly fusing these conflicting results may lead to ghosting artifacts, where structural details appear blurry or inconsistent (``Fused without CVA" in Figure~\ref{fig:CVA}).

To resolve these conflicting structural details, Cross-View Alignment (CVA) enforces feature consistency by using the front-view trajectory $X^{1:N}_{\text{front}}$ as a reference to guide the denoising process of the anchor-view trajectory $X^{1:N}_{\text{anchor}}$. Specifically, CVA injects intermediate features from the front-view denoising process into the anchor-view denoising process using a cross-view feature attention mechanism.

During each denoising step $t$, we modify the self-attention layers of the diffusion network by injecting key-value pairs from the front-view trajectory into the attention computation of the anchor-view trajectory. The feature-aligned attention output $\mathcal{A}_t^{\text{CVA}}$ is computed as: 
\begin{equation}\label{eq:6} 
\mathcal{A}_{t}^{\text{CVA}} = \text{softmax} \left(\frac{Q_{t}^{a} \cdot (K_{t}^{f} \oplus K_{t}^{a})^T}{\sqrt{d_k}}\right) \cdot (V_{t}^{f} \oplus V_{t}^{a}),
\end{equation}
where $\oplus$ denotes the concatenation operation, $d_k$ is the embedding dimension, $Q_{t}^{a}$, $K_{t}^{a/f}$, and $V_{t}^{a/f}$ denote the query, key, and value features from the V3DM along the respective anchor-view (a) or front-view (f) camera trajectory.
By injecting front-view priors into the denoising process, CVA guides the model to maintain consistent structure across views, preventing ghosting artifacts while improving geometric fidelity (``Fused with CVA" in Figure~\ref{fig:CVA}).

%% file: sec/4_experiments.tex
\begin{figure*}
  \centering
  \includegraphics[width=\textwidth]{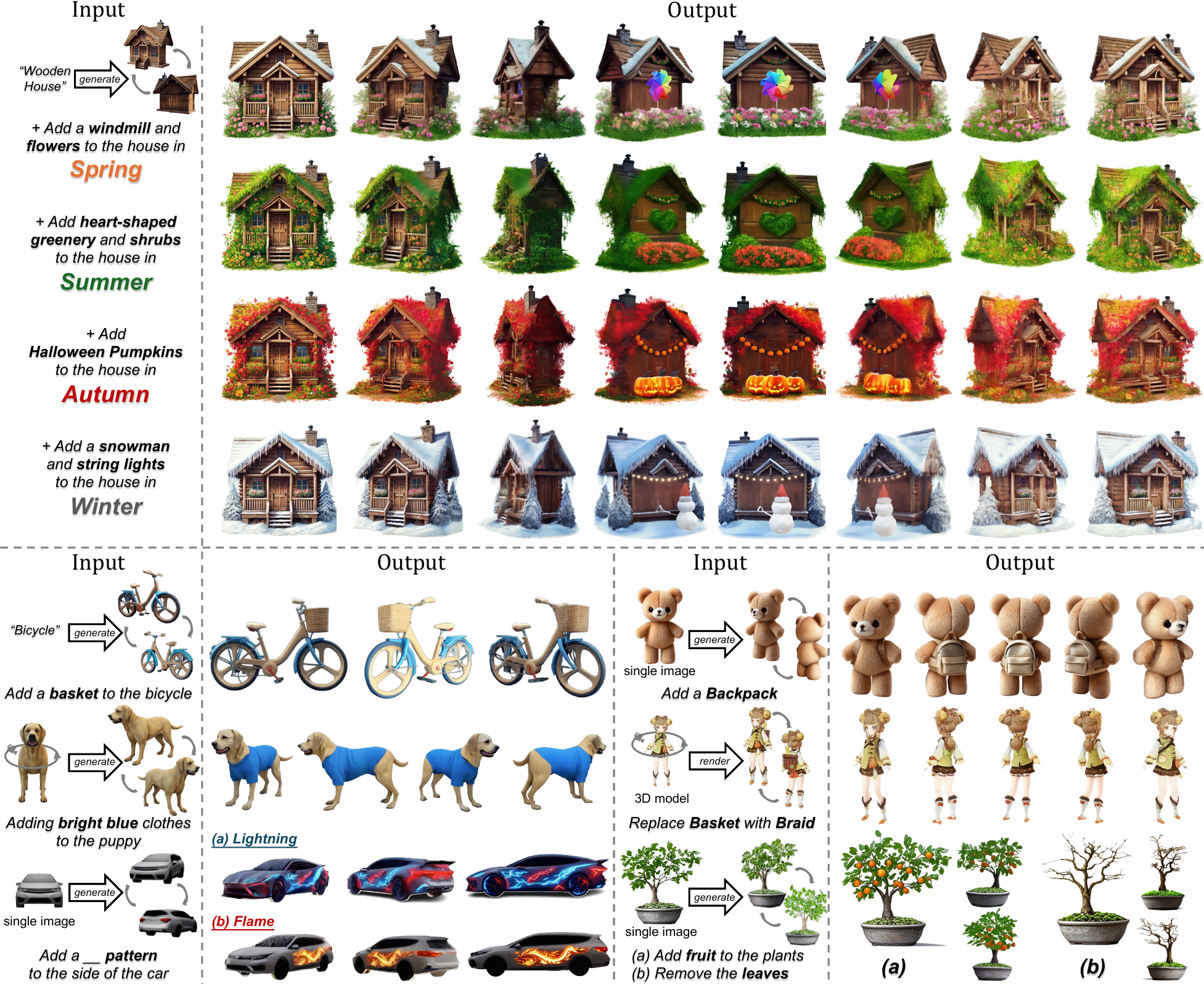}
  \vspace{-5mm}
  \caption{\textbf{Examples of local element editing with Edit360.} Our framework accepts various inputs (text, single images, or 3D models) and editing instructions, showing precise local modifications including element insertion, deletion, and replacement across various subjects.}
  \vspace{-0mm}
  \label{fig:EditResult}
\end{figure*}

\begin{figure*}
  \centering
  \includegraphics[width=\textwidth]{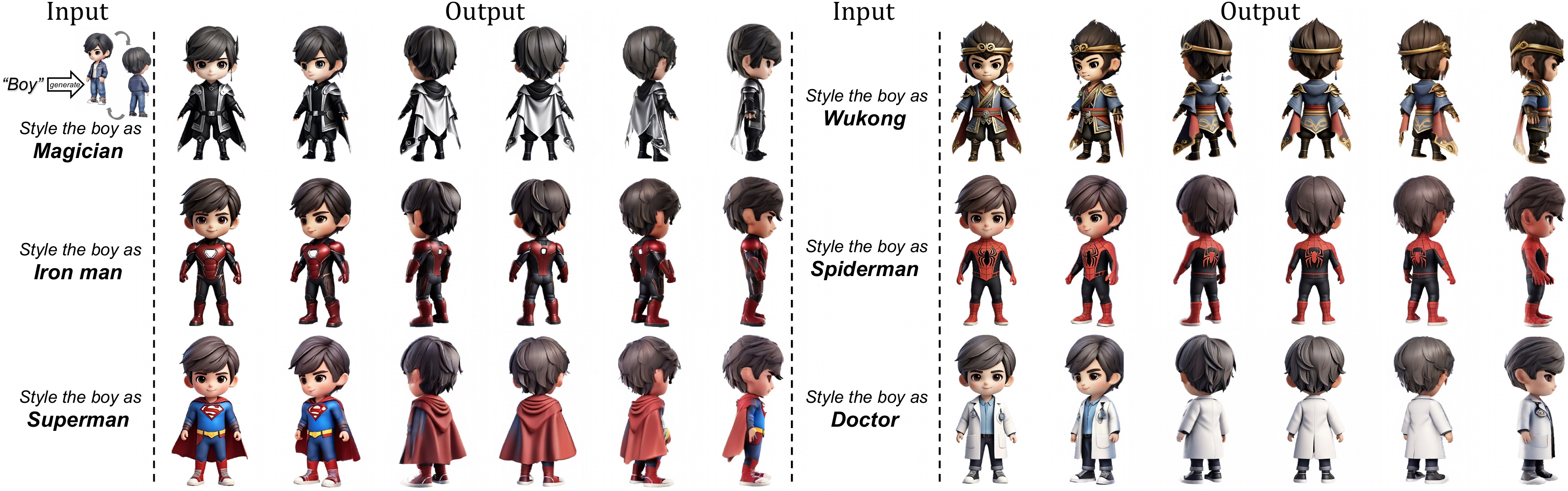}
  \vspace{-5mm}
  \caption{\textbf{Examples of global style transfer with Edit360.} Our method successfully transfers diverse character styles (Magician, Iron Man, Superman, Wukong, Spiderman, and Doctor) while preserving the original character's identity.}
  \vspace{-4mm}
  \label{fig:StyleResult}
\end{figure*}

\section{Experiments}
\begin{figure*}
  \centering
  \includegraphics[width=0.99\textwidth]{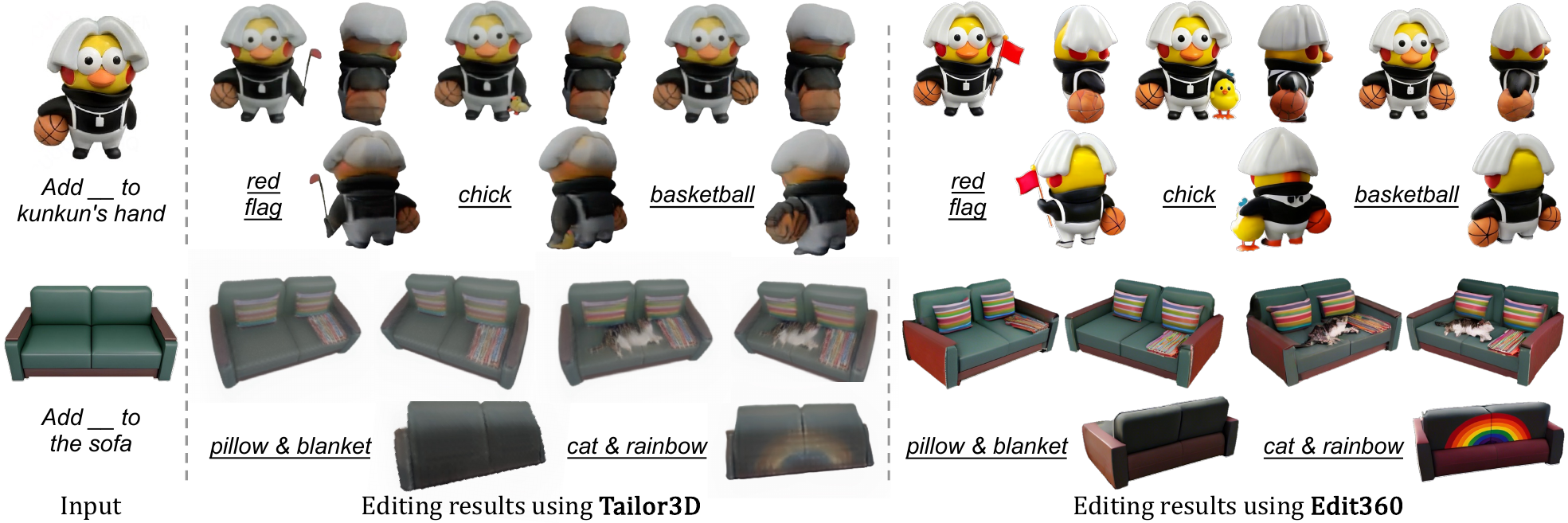}
  \vspace{-2.5mm}
  \caption{
  Qualitative comparison of editing results between Tailor3D~\cite{voleti2024sv3d} and our Edit360 framework. Edit360 produces sharper, higher-resolution textures and richer geometric details, resulting in more realistic 3D editing results.}
  \vspace{-3mm}
  \label{fig:tailor3d}
\end{figure*}

\begin{figure*}
  \centering
  \includegraphics[width=0.99\textwidth]{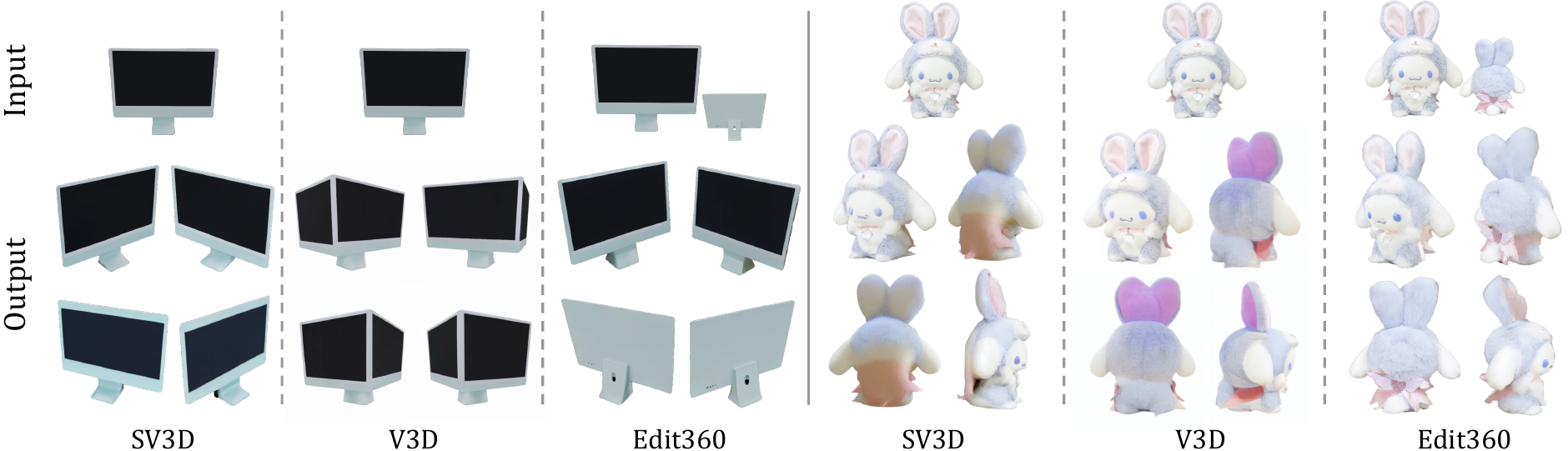}
  \vspace{-2.5mm}
  \caption{
  Qualitative comparison between existing V3DMs and our enhanced Edit360 framework. Our method surpasses both SV3D~\cite{voleti2024sv3d} and V3D~\cite{chen2024v3d} by avoiding multi-faced Janus problems (e.g., the monitor) and generating richer back details (e.g., the doll's rear appearance).
  }
  \vspace{-4.5mm}
  \label{fig:GenResult}
\end{figure*}

To demonstrate the effectiveness of our proposed Edit360 framework, we conduct both qualitative and quantitative experiments across 3D editing and generation tasks. Additionally, we perform ablation studies to assess the impact of key components in the Edit360.

\subsection{Experimental Setup}\label{sec:exp_setup}
\textbf{Combined with Different V3DMs.}
Several video 3D diffusion models (V3DMs)~\cite{chen2024v3d, voleti2024sv3d, pang2024envision3d, han2025vfusion3d} fine-tuned on the Objaverse dataset\cite{deitke2022objaverseuniverseannotated3d} can generate dense views from a single front-view input for 3D reconstruction. We evaluate our method on two representative models: $SV3D^u$~\cite{voleti2024sv3d} generating 21 horizontal views (576×576) and $V3D$~\cite{chen2024v3d} producing 18 views (512×512), both at 0-degree elevation. Edit360 extends these V3DMs to support multiple input views without additional training, with all inference performed on a single NVIDIA RTX 4090 GPU.

\vspace{1mm}
\noindent
\noindent \textbf{Datasets and Metrics.}
For 3D editing tasks, we demonstrated our model's fine-grained editing capabilities through both qualitative comparisons and quantitative user studies on diverse data sources, including single images (either generated by text-to-image models~\cite{10140348} or captured from the real world) and existing 3D models. For novel multi-view synthesis evaluation, we quantitatively assessed our method's ability to generate novel views using the unseen Google Scanned Objects (GSO)\cite{Downs2022GoogleSO} and OmniObject3D~\cite{wu2023omniobject3d} datasets. Given the large number of similar objects with minor color variations existed in GSO dataset, we followed SV3D\cite{voleti2024sv3d}, filtering out 300 objects with significant shape differences from the dataset to ensure diversity and minimize redundancy in our evaluation. We generate dense orbit videos based on camera trajectories and compare each generated frame to the corresponding ground truth using multiple metrics: LPIPS~\cite{8578166}, PSNR, and SSIM~\cite{1284395} for comprehensive evaluation. 

\subsection{3D Asset Editing}
Our framework enables precise local editing of diverse 3D assets from arbitrary viewpoints, supporting operations such as element insertion, replacement, and removal. Moreover, it facilitates comprehensive style transformations through specified editing directives.

\vspace{1mm}
\noindent\textbf{Local Element Editing.}
Figure~\ref{fig:EditResult} demonstrates our framework's capability for localized editing through element-level modifications. In the first example, we input a front view of a cabin generated by DALL·E using the prompt ``wooden house" and applied seasonal edits with to create 360-degree view videos. For each season, we add specific elements: a windmill and flowers for spring, heart-shaped greenery and shrubs for summer, Halloween pumpkins for autumn, and both a snowman and string lights for winter. Additionally, we demonstrate the framework's versatility across various subjects—characters, vehicles, animals, and plants—through precise modifications including accessory addition, component replacement, and element removal.

\vspace{0.5mm}
\noindent\textbf{Global Style Transfer.}
Edit360 enables global style transfer across 360-degree views by replacing editing instructions with style-specific prompts, while preserving the subject's identity. As shown in Figure~\ref{fig:StyleResult}, starting from a text-generated boy image, we successfully transform his appearance into various fantasy and superhero characters (Magician, Iron Man, Superman, Wukong, Spiderman, and Doctor) while preserving his distinct facial features.

\begin{table}
  \centering
  \resizebox{0.75\linewidth}{!}{
  \begin{tabular}{lccc}
    \toprule
    Approach & Geometric & Texture & Overall \\
    \midrule
 Tailor3D& 3.32& 2.77&3.02\\
 Edit360 (Ours)& 4.56& 4.49&4.52\\
 \bottomrule
  \end{tabular}}
  \vspace{-2mm}
  \caption{Comparison of user study ratings between Edit360 and Tailor3D for 3D editing tasks across geometric consistency, texture preservation, and overall performance.}
  \vspace{-1mm}
  \label{tab:user_study}
\end{table}

\vspace{0.5mm}
\noindent\textbf{Comparison in 3D editing task.} 
Tailor3D is a concurrent work that explores extending 2D editing to 3D domain. However, they are constrained by fixed editing viewpoints (front and back views), making it challenging to edit other angles (e.g., side views). Edit360 supports flexible editing from any viewpoint, and achieves better editing results by selecting anchor views for specific editing tasks. Furthermore, by leveraging rich visual priors from video diffusion models, our method achieves more refined and realistic editing effects. We conduct qualitative comparisons using the same examples shown in their paper, as demonstrated in Figure~\ref{fig:tailor3d}. Edit360 achieves superior geometric structure and richer texture details. Additionally, we conducted a user study to evaluate 3D editing quality, recruiting 50 participants to assess eight editing examples (each presented from four distinct viewpoints). As illustrated in Table~\ref{tab:user_study}, evaluators rated our method on a 5-point Likert scale across three dimensions: geometric consistency, texture detail, and overall performance. The results clearly demonstrate the superiority of our approach in these critical aspects.

\begin{table}
  \centering
  \begin{adjustbox}{width=\linewidth}
    % \begin{tabular}{@{}lcccccc@{}}
    \begin{tabular}{lcccccc}
      \toprule
      \multirow{2}{*}{Model} & \multicolumn{3}{c}{GSO Dataset} & \multicolumn{3}{c}{OmniObject3D Dataset} \\
      & \multicolumn{1}{c}{LPIPS $\downarrow$} & \multicolumn{1}{c}{PSNR $\uparrow$} & \multicolumn{1}{c}{SSIM $\uparrow$} & \multicolumn{1}{c}{LPIPS $\downarrow$} & \multicolumn{1}{c}{PSNR $\uparrow$} & \multicolumn{1}{c}{SSIM $\uparrow$} \\
      \midrule
      Zero123~\cite{liu2023zero}                     & 0.13 & 17.29 & 0.79 & 0.17 & 15.50 & 0.76 \\
      Zero123XL~\cite{Deitke2023ObjaverseXLAU}        & 0.14 & 17.11 & 0.78 & 0.18 & 15.36 & 0.75 \\
      Stable Zero123~\cite{stabilityAI2023stablezero123}  & 0.13 & 18.34 & 0.78 & 0.15 & 16.86 & 0.77 \\
      Free3D~\cite{zheng2024free3dconsistentnovelview} & 0.15 & 16.18 & 0.79 & 0.16 & 15.29 & 0.78 \\
      EscherNet~\cite{Kong2024EscherNetAG}           & 0.13 & 16.73 & 0.79 & 0.17 & 14.63 & 0.74 \\
      SV3D~\cite{voleti2024sv3d}                     & 0.09 & 21.14 & 0.87 & 0.10 & 19.68 & 0.86 \\
      \midrule
      Edit-V3D ($v^0$)& 0.09 & 21.30 & 0.91&  0.10&  19.64& 0.87\\
      Edit-SV3D ($v^0$)&      0.08 &       21.65 &          0.86 &  0.10&  19.77& 0.88\\
 Edit-V3D ($v^{0}$ \& $v^{i}$)& 0.09& 21.42& \textbf{0.92}& 0.09& 19.98&0.87\\
 Edit-SV3D ($v^{0}$ \& $v^{i}$)& \textbf{0.07}& \textbf{22.17}& 0.90& \textbf{0.08}& \textbf{20.44}&\textbf{0.89}\\ 
 \midrule \rowcolor{gray!20}
      Edit-SV3D (2GT input)& 0.06& 26.32&          0.93& 0.07&  24.52& 0.91\\
      \bottomrule
    \end{tabular}
  \end{adjustbox}
  \vspace{-2mm}
  \caption{Quantitative comparison of our Edit360 framework with recent novel view synthesis methods.}
  \vspace{-4mm}
  \label{tab:GenResult}
\end{table}

\subsection{Novel Multi-View Synthesis}
\label{subsec:GenResult}
While Edit360 primarily focuses on 3D editing capabilities, it's crucial to validate that our framework preserves multi-view consistency when introducing additional view conditions during the editing process. Therefore, we evaluate Edit360's performance in novel view synthesis tasks, demonstrating that our framework do not compromise the underlying generation capabilities of base V3DM models.
As shown in Table~\ref{tab:GenResult}, we quantitatively compare with some state-of-the-art novel view synthesis methods and two V3DM methods (V3D~\cite{chen2024v3d}, SV3D~\cite{voleti2024sv3d}), as well as the V3DM methods enhanced with Edit360. For fair comparison, all the single-view input methods utilize a ground-truth front view $v^0$ as input, an additional anchor view $v^{i}$ for Edit360 is generated using the original single front-view input V3DM model (where $i$ is a randomly selected frame index from the dense-view sequence).

Additionally, we evaluate Edit360 with a two-view ground-truth input setting (``2GT input" in Table~\ref{tab:GenResult}), where two ground-truth views are provided as input anchor views. This setting reaches a significant performance improvement, highlighting the superior upperbound of our framework when leveraging multi-view ground-truth inputs. 
The qualitative comparison is shown in Figure~\ref{fig:GenResult}. We observe that conventional V3DM methods (SV3D~\cite{voleti2024sv3d}, V3D~\cite{chen2024v3d}) often generate incorrect or overly smooth back views, failing to preserve the complex textures and details present in the front view (e.g., the doll in Figure~\ref{fig:GenResult}). Additionally, these methods sometimes suffer from the ``Janus problem", where the front-view content is incorrectly repeated in other viewpoints (e.g., the monitor in Figure~\ref{fig:GenResult}). In contrast, our framework ensures texture consistency across surrounding views by leveraging multi-view inputs (\textit{e.g.}, front and back views). 

\begin{table}
\centering
\resizebox{\columnwidth}{!}{
\begin{tabular}{cc|ccccc}
\toprule
 SPF& CVA& LPIPS $\downarrow$ & PSNR $\uparrow$ & SSIM $\uparrow$ & CLIP-S $\uparrow$ &MSE $\downarrow$ \\
\midrule
 &  &  0.11&  20.57&   0.83& 0.85&0.02\\
\cmark&  & 0.08& 20.90& 0.87& 0.85&0.02\\
\cmark& \cmark &  \textbf{0.07}&  \textbf{22.17}&  \textbf{0.90}& \textbf{0.88}&0.02\\
\bottomrule
\end{tabular}
}
\vspace{-2mm}
\caption{Ablation study on key components of the Edit360: Spatial Progressive Fusion (SPF) and Cross-View Alignment (CVA), demonstrating their individual contributions to performance.}
\vspace{-2mm}
\label{tab:ablation}
\end{table}

\begin{figure}
\centering
\includegraphics[width=\linewidth]{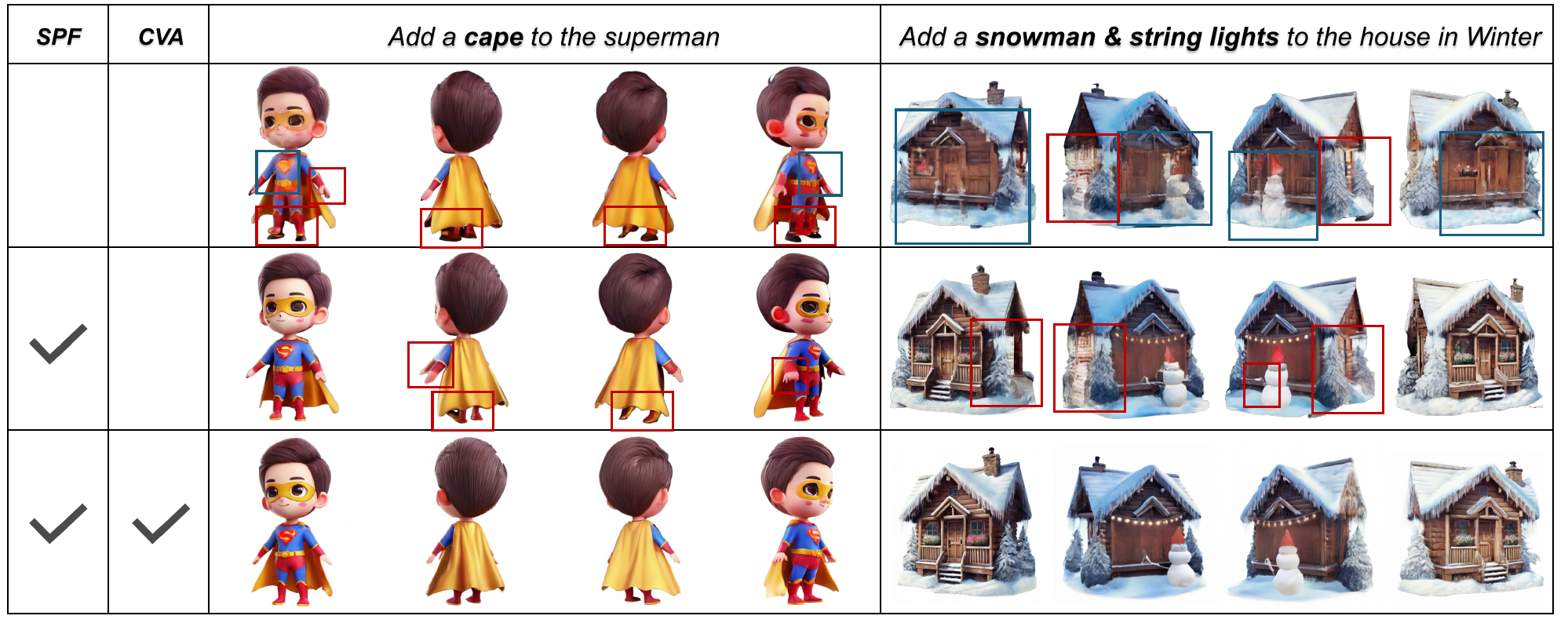}
\vspace{-6mm}
\caption{Visual ablations on SPF and CVA:
The first row shows results by directly adding features after rotation alignment (Eq.~4). Red boxes indicate inconsistencies, while blue boxes indicate over-smoothing artifacts. With both modules, our method (third row) produces more consistent and detailed results.}
\vspace{-5mm}
\label{fig:VA}
\end{figure}

\subsection{Ablation Studies}
We present the following ablation studies on the GSO dataset based on the experimental settings outlined in Sec \ref{sec:exp_setup}.
As shown in Table~\ref{tab:ablation}, we evaluate the impact of two key components: Spatial Progressive Fusion (SPF), and Cross-View Alignment (CVA). The baseline model uses simple linear interpolation with equal weights across all frames. The results highlight the importance of each component in our framework. Additionally, Figure~\ref{fig:CVA} and Figure~\ref{fig:VA} provide visual ablations to demonstrates the effect of the proposed SPF and CVA modules. 

% Further ablation experiments examining spatial weights and multi-scale fusion in SPF, as well as the impact of anchor view angles, are presented in the supplementary materials.

%% file: sec/5_conclusion.tex
\section{Conclusions}
In conclusion, Edit360 offers a novel training-free solution for 3D asset editing, enabling precise 360-degree customization based on user-specified instructions. By identifying the optimal editing views for specific editing tasks, and consistently integrating 2D edits into the entire 3D asset through Spatial Progressive Fusion (SPF) and Cross-View Alignment (CVA), we achieve precise 3D editing. This approach overcomes the limitations of prior methods, facilitating complex editing tasks while maintaining spatial coherence and detail throughout the entire 3D model. With its capacity for detailed and flexible 3D customizations, Edit360 shows great potential for applications in animation, gaming, and virtual reality requiring 3D asset manipulation.